\documentclass[aps,pre,preprint,floatfix]{revtex4}
\usepackage{graphicx}
\usepackage{amsmath}
\usepackage{amsfonts}
\usepackage{amssymb}
\usepackage{graphicx}
\usepackage{bm}%
\usepackage{verbatim}

\usepackage{color}

\usepackage{algorithm}
\usepackage{algpseudocode}

\newcommand{\bq}{\begin{equation}}
\newcommand{\eq}{\end{equation}}

\begin{document}
\title{A backward Monte-Carlo method for time-dependent \\ runaway electron simulations\footnote{{\bf NOTICE OF COPYRIGHT}: This manuscript has been authored by UT-Battelle, LLC under Contract No. DE-AC05-00OR22725 with the U.S. Department of Energy. The United States Government retains and the publisher, by accepting the article for publication, acknowledges that the United States Government retains a non-exclusive, paid-up, irrevocable, worldwide license to publish or reproduce the published form of this manuscript, or allow others to do so, for United States Government purposes. The Department of Energy will provide public access to these results of federally sponsored research in accordance with the DOE Public Access Plan (http://energy.gov/downloads/doe-public-access-plan).}}
\author{Guannan Zhang}
\author{Diego del-Castillo-Negrete}
\affiliation{Oak Ridge National Laboratory \\ Oak Ridge, Tennessee 37831-8071, USA}

\begin{abstract}
Kinetic descriptions of runaway electrons (RE) are usually based on   Fokker-Planck models that determine the probability distribution function (PDF) of RE in 2-dimensional momentum space. Despite of the simplification involved, the Fokker-Planck equation can rarely be solved analytically and direct numerical approaches (e.g., continuum and particle-based Monte Carlo (MC)) can be time consuming, especially in the computation of asymptotic-type observables including the runaway probability, the slowing-down and runaway mean times, and the energy limit probability. Here we present a novel backward MC approach to these problems based on backward stochastic differential equations (BSDEs) that  describe the dynamics of the runaway probability by means of the  Feynman-Kac theory. The key ingredient of the backward MC algorithm is to place all the particles in a runaway state and simulate them backward from the terminal time to the initial time. As such, our approach can provide much faster convergence than direct MC methods (by significantly reducing the number of particles 
required to achieve a prescribed accuracy) while at the same time maintaining the 
advantages of particle-based methods (compared to continuum approaches). 
The proposed algorithm is unconditionally stable, can be parallelized as easy as the direct MC method, 
and its extension to  dimensions higher than two is straightforward, 
thus paving  the way for conducting large-scale RE simulation.
\end{abstract}
\maketitle

\section{Introduction}
  
At high enough velocities, the drag force on a particle due to Coulomb collisions in a plasma decreases as the particle velocity increases. As a result, in the presence of a strong enough parallel electric field, fast electrons can ``runaway" and be continuously accelerated
 \cite{dreicer_1959,gurevich_1961,connor_hastie_1975}. In magnetically confined fusion plasmas runaway electrons (RE) can be generated during magnetic, disruptions due to the strong electric field resulting from the rapid cooling of the plasma,
see for example  Refs.~\cite{gill_etal_2000,hollman_etal_2013}. Understanding this phenomena has been an area of significant interest because of the  potential impact that RE can have to the safe operation of ITER. In particular, if not avoided or mitigated, RE can severely damage plasma facing components, see for example  Refs.~\cite{Hender_2007,Hollmann_etal_2011}. 

The goal of this paper is to present a novel application of a backward Monte-Carlo method to the study of RE in phase space. 
Although this method can be applied to problems of  arbitrary dimensions, here we focus on RE in 
$2$-dimensional phase space with coordinates $(p, \theta)$ where $p$ denotes the magnitude of the relativistic momentum and $\theta$ the 
pitch angle, i.e. the angle between the electron's velocity and the magnetic field. As it is well known, in this case the dynamics 
of the RE probability distribution function, $f(p,\theta,t)$, is determined by the 
Fokker-Planck (FP) equation describing the competition between the electric field acceleration, Coulomb collisions, synchrotron radiation damping, and sources describing the  secondary generation of RE due to head-on collisions. 
Although approximate, the $2$-D Fokker-Planck model provides important physical  insights and, as a result, efforts have been devoted to the development of efficient and accurate  methods of its solution. At the same time, because of its relative simplicity
(compared higher dimensional problems) the $2$-D Fokker-Planck model is an excellent testbed for the development and testing of novel 
numerical schemes.  

The standard Fokker-Planck problem pertains the solution of the forward evolution of the probability distribution function, $f(p,\theta,t>t_0)$, 
for a given initial condition, $f(p,\theta,t=t_0)$. This problem can be solved either using continuum methods (e.g., finite-difference, finite-elements, or spectral) or using particle-based Monte-Carlo methods. However, in the study of RE, as well as several other applications, 
important questions involve statistical observables different to $f(p,\theta,t)$. Examples of particular interest to the present paper are 
the probability, $P_{\rm RE}(t,p,\theta)$, that an electron with phase space coordinates $(p, \theta)$ will runaway on, or before, a time $t$,
the expected runaway time $T_{\rm RE}$, the expected loss time $T_{\rm Loss}$, and the RE production rate. 
These problems can be difficult to address from the solution of the forward evolution of the Fokker-Planck equation because it is not clear a priori what initial conditions will give the  desired final   state, in this case the runaway state. 
In the general theory of Fokker-Planck equations, see for example Secs.~3.6 and 5.2 in Ref.~\cite{Gardiner_2004}, the alternative for this type of problems is to focus on the  so-called backward  Fokker-Planck equation that determines the evolution of $f(p,\theta,t<t_f)$ given the {\em final} state $f(p,\theta,t=t_f)$. 
From a mathematical point of view, the backward Fokker-Planck equation is the adjoint of the standard, forward Fokker-Planck equation. 

The  method we are proposing is different from those based on the solution of the adjoint Fokker-Planck equation, e.g. Ref.~\cite{karney_1986,liu_etal_2016,liu_etal_2017}, and also different from the naive  ``brute force" approach based on direct Monte-Carlo simulations.  Methods based on the solution of the adjoint Fokker-Planck equation can potentially face the well-known generic shortcomings of the solution of partial differential equations including the unfavorable scaling with increasing dimensionality, the need to use properly chosen grids and/or special basis functions to accommodate the geometry of the integration domain, and stability issues when considering time-dependent problems. Monte-Carlo methods on the other hand face the need to include an extremely large number of particles to reduce the statistical noise. Our approach is based on the Feynmann-Kac formula relating the solution of the adjoint Fokker-Planck equation and the corresponding system of stochastic differential equations. As
it will be explained in the next section, this method offers significant advantages over  
partial differential equation methods for solution of the backward Fokker-Planck  and direct Monte-Carlo methods. 
These advantages are particularly relevant in the study of the time dependent evolution of the runaway probability, which, going beyond previous studies limiting attention to steady state regimes  \cite{liu_etal_2016,liu_etal_2017}, is one of the goals of the present paper. 

The rest of the paper is organized as follows. In the next section we present the $2-D$ phase space runaway electron model in the continuum 
(partial differential equations) Fokker-Planck  version and in the equivalent particle-based (stochastic differential equations) Langevin formulations.  
We also formulate the problem of interest and its connection with the adjoint Fokker-Planck equation. 
Section~\ref{BMC} discusses the mathematical foundation and the numerical algorithm of the  proposed Backward Monte-Carlo (BMC) method.
Section~\ref{applications} discusses applications of the   BMC to problems of physical interest including the computation of the time-dependent probability of runaway, the expected runaway and loss times, and the production rate of runaway electrons. A summary of results and concluding remarks are presented in Section \ref{conclusions}.

\section{Model and problem formulation}
\label{model}

The Backward Monte Carlo (BMC) method can in principle be applied to problems of arbitrary dimension. However, 
 in order to illustrate the method in the simplest possible setting in this paper we 
 neglect spatial-effects and limit attention to 2-dimensional models describing the  probability density function, $f$, of RE in the $(p,\xi)$ phase space where $p$ denotes the magnitude 
of the relativistic momentum ${\bf p}=m_e \gamma {\bf v}$, with $m_e$ the electron mass and 
$\gamma=1/\sqrt{1-v^2/c^2}$ the relativistic factor, and $\xi=\cos \theta$, where $\theta$ is the pitch angle, i.e. $\xi={\bf v}\cdot {\bf B}/vB$. 
 The extension of the method to higher dimensions will be discussed at the end of Sec.~\ref{BMC}. 
 
The dynamics of $f(t,p,\xi)$ is assumed to be determined by the  
 test particle relativistic Fokker-Planck equation \cite{anderson_etal_2001,stahl_etal_2015}
\bq
\label{FP}
\frac{\partial f}{\partial t}= {\cal F} + {\cal C} +{\cal R} \, ,
\eq
where 
\bq
{\cal F}=- E\left[ \frac{\xi}{p^2}\frac{\partial}{\partial p}\left(p^2 f\right)+\frac{\partial}{\partial \xi} \left( \frac{1-\xi^2}{p} f\right)\right]
\eq
is the acceleration due to the electric field, $E$,  which neglecting self-consistent effects is  assumed constant,  
\bq
\label{col}
{\cal C}=\frac{1}{p^2} \frac{\partial}{\partial p} \left[ \left(1+p^2\right) f\right]
+ \frac{\nu_c}{2} \frac{\partial}{\partial \xi} \left[ \left( 1 - \xi^2\right) \frac{\partial f}{\partial \xi} \right]
\eq
is the  small-angle Coulomb collision operator   in the relativistic limit with $\nu_c = \left(Z+1\right) \sqrt{1+p^2}/p^3$ and 
$Z$ denoting the ion effective charge,
and 
${\cal R}$ is the synchrotron radiation reaction force given by
\bq
\label{rr}
{\cal R}= \frac{1}{\tau} \left\{
\frac{1}{p^2} \frac{\partial}{\partial p} \left[ p^3 \gamma \left(1-\xi^2\right) f
\right ] - \frac{\partial}{\partial \xi} \left[ \frac{1}{\gamma} \xi \left(1-\xi^2\right) f
\right]
\right \} \, .
\eq
In writing the model we have used dimensionless variables. The momentum $p$ has been normalized using $m_e c$,
where $c$ denotes the speed of light, the electric field has been normalized using the Connor-Hastie critical electric field,
$E_c=n_e e^3 {\rm ln} \Lambda/(4 \pi \epsilon_0^2 m_e c^2)$, where $\Lambda$ is the Coulomb logarithm, 
and the time has been normalized using the relativistic collision time scale, $\tau_{c}=m_e c/(E_c e)$. The parameter
${\tau}=\tau_r/\tau_c$ where $\tau_r=6 \pi \epsilon_0 m_e^3 c^3/(e^4 B^2)$
is the synchrotron  radiation time scale, with $e$  the electron charge, $B$ the characteristic magnitude of the magnetic field, and $\epsilon_0$ is the vacuum permittivity. In these variables, $\gamma=\sqrt{1+p^2}$.

From the connection between Fokker-Planck equations and stochastic differential equations (SDEs), 
see for example Sec.~4.3 in Ref.~\cite{Gardiner_2004}, it follows that the SDEs corresponding to 
Eq.~(\ref{FP}) are
%
\begin{equation}\label{e1}
\begin{aligned}
d p_t & =b_1(p_t, \xi_t)\, dt, \\
d \xi_t & = b_2(p_t, \xi_t)\, dt + \sigma_2 (p_t,\xi_t)\, dW_t,
\end{aligned}
\end{equation}
where, in this case, the drift coefficients $b_1, b_2$ and the diffusion coefficient $\sigma_2$ are given by
\begin{equation}\label{e2}
\begin{aligned}
& b_1 = E \xi - \frac{\gamma p}{\tau}  \left( 1- \xi^2 \right)-\frac{1+p^2}{p^2},\\
& b_2 =  \frac{E \left(1-\xi^2\right)}{p} + \frac{\xi \left(1-\xi^2\right)}{\tau \gamma} -\xi \nu_c,\\
&  \sigma_2 =  \sqrt{ \nu_c \left(1-\xi^2\right)}, \,
\end{aligned}
\end{equation}
and $W_t$ is the standard Brownian motion (Wiener process) according to which the increments $d W_t$ are drawn from a Gaussian distribution with
zero mean and variance equal to $dt$.
%
%

The  problem we want to address is the computation of the probability that an electron with coordinates $(p,\xi)$ will runaway at, or before, a prescribed time.  By ``runaway" we mean that, as a result of the electric field acceleration, the electron will reach a 
prescribed momentum, $p_*$. It is important to keep in mind that there is some freedom in the definition of when a give electron is labeled as runaway because  
  $p_*$  is a free parameter. The dependence of the runaway probability on $p_*$ becomes negligible for large enough $p_*$, which is the reason why this dependence is not usually accounted for explicitly. 
More formally, for a given 
$(t,p, \xi) \in [0,T] \times  [p_{\min}, p_*] \times [-1,1] $, where $p_{\min}$ is a lower momentum boundary,
the runaway probability, $P_{\rm RE}(t, p, \xi)$, is defined as 
the probability that an electron located at  $(p,\xi)$ at the initial time instant $t_0 = 0$ will acquire a momentum $p_*$ on, or before, $t >0$. 

Note that the runaway probability, $P_{\rm RE}(t,p,\xi)$, is different from the solution, $f(t,p,\xi)$, of the Fokker-Planck equation in Eq.~\eqref{FP},
that gives the probability for an electron to be at $(p,\xi)$ at time $t$. 
Given $f(t,p,\xi)$, 
$P_{\rm RE}(t,p,\xi)$ can be obtained from the conditional expectation 
\begin{equation}\label{REP}
P_{\rm RE}(t,p,\xi) = \mathbb{E}[\chi(p_t, \xi_t) \,|\, p_0 = p, \xi_0 = \xi] = \int_{\mathbb{R}^2} \chi(p_t, \xi_t) f(t,p_t,\xi_t\, |\, p, \xi)\, dp_t\, d\xi_t,
\end{equation}
where $(p_t, \xi_t)$ is the solution of the SDEs in Eq.~\eqref{e1} with the initial condition $(p_0,\xi_0) = (p,\xi)$, $f(t,p_t,\xi_t \,|\, p, \xi)$ is the solution of Eq.~\eqref{FP} with the initial condition $f(0,p,\xi)=\delta (p-p_0) \delta (\xi-\xi_0)$
 and  $\chi(p_t, \xi_t)$ is defined as
\begin{equation}\label{chafun}
\chi(p_t,\xi_t) = 
\left\{
\begin{aligned}
&1, \;\; \text{ if } \; p_t \ge p_{*},\\
&0, \;\; \text{ otherwise},
\end{aligned}
\right.
\end{equation}
which indicates whether a realization $(p_t, \xi_t)$ of the SDEs is a runaway path. 

A straightforward way to compute $P_{\rm RE}$ at a given point, $(p,\xi)$, in phase space at a given time $t$ 
is to use the forward, ``brute-force", MC method. That is, to simulate a very large number of paths, $(p_t,\xi_t)$,
by solving the SDEs in Eq.~\eqref{e1}, with initial condition $(p_0,\xi_0)=(p,\xi)$, and substitute the obtained paths in Eq.~\eqref{REP} to approximate the expectation. 
Despite its simplicity, this direct method is very inefficient due to the slow convergence of the MC sampling, and the fact that   a new set of paths has to be generated to compute $P_{\rm RE}$ at each point in phase space. 

An alternative to the direct Monte-Carlo method is based on the use of the backward Fokker-Planck equation that considers the evolution of 
%
\begin{equation}\label{PPP}
P(t,p,\xi) = P_{\rm RE}(T-t, p, \xi) 
\end{equation}
for 
$(t,p,\xi) \in [0,T] \times  [p_{\min}, p_*] \times [-1,1]$. 
The function  $P(t,p,\xi)$ is given by the  solution of the terminal value problem
\begin{equation}\label{KBE}
\left\{
\begin{aligned}
& \frac{\partial P}{\partial t} +  b_1 \frac{\partial P}{\partial p} + b_2 \frac{\partial P}{\partial \xi} + \frac{ \sigma_2^2}{2}\frac{\partial^2 P}{\partial \xi^2} = 0,\\
& P(T,p,\xi) = \chi(p,\xi),
\end{aligned}\right.
\end{equation}
where $b_1, b_2,  \sigma_2$ are given in Eq.~\eqref{e2}, and $\chi(p,\xi)$ is defined in Eq.~\eqref{chafun}. 
Formally, Eq.~\eqref{KBE} is the adjoint problem of the Fokker-Planck equation in \eqref{FP}, see Ref.~\cite{Gardiner_2004} for details.
%
%
%
%
%

The adjoint method has found applicability in a wide range of problems. An early application in the context of runaway electrons is 
Ref.~\cite{karney_1986}. More recently, in Refs.~\cite{liu_etal_2016,liu_etal_2017}  the 
time-independent adjoint Fokker-Planck equation was used to study the steady-state runaway probability function and the expected loss times for highly relativistic runaway electrons. 
In the following section we discuss a novel application of the Backward Monte Carlo Method for an efficient and accurately solution of the  time dependent  evolution of the runaway probability function and the expected loss times. 

%

\section{The Backward Monte Carlo Method}
\label{BMC}

The theoretical foundation of the backward Monte Carlo method is the Feynman-Kac theory that links the SDE in Eq.~\eqref{e1}-\eqref{e2} to the time-dependent adjoint equation in Eq.~\eqref{KBE}. In particular, according to Feynman-Kac formula \cite{Karatzas}, the solution $P(t,p,\xi)$ of the adjoint equation (\ref{KBE}) is given by the conditional expectation:
\begin{equation}\label{e5}
\begin{aligned}
P(t,p,\xi) & =  \mathbb{E}[P(T, p_{T}, \xi_{T}) \,|\, p_t = p, \xi_t = \xi],\\
\end{aligned}
\end{equation}
where $P(T, p_{T}, \xi_{T}) = \chi(p_T,\xi_T)$. 
%
%
As opposed to solving the adjoint partial differential equation in ~\eqref{KBE}, the key idea of the BMC method is to 
solve Eq.~(\ref{e5}) directly. 
%
%

The first step of the proposed numerical method is to introduce a uniform partition of the time interval $[0,T]$, 
\begin{equation}\label{e9}
\mathcal{T} = \{0 = t_0 < t_1 < \cdots < t_N = T\},
\end{equation}
where $\Delta t = t_n - t_{n-1}$. 
Due to the Markovian property of the SDE in Eq.~(\ref{e1}),  Eq.~(\ref{e5}) implies that, within the time interval $[t_n, t_{n+1}]$,
\begin{equation}\label{e7}
P(t_n, p, \xi) = \mathbb{E} \left[ P(t_{n+1}, p_{t_{n+1}}, \xi_{t_{n+1}}) \;|\; p_{t_n}= p, \xi_{t_n} = \xi \right].
\end{equation}
For sufficiently small $\Delta t$, $(p_{t_{n+1}}, \xi_{t_{n+1}})$ can be obtained from 
$(p_{t_{n}}, \xi_{t_{n}})$ by approximating the SDE in Eq.~(\ref{e1}) using a simple Euler scheme, i.e.,
\begin{equation}\label{e6}
\begin{aligned}
p_{t_{n+1}} & = p + b_1(p, \xi)\, \Delta t, \\
\xi_{t_{n+1}} & =  \xi + b_2(p, \xi)\, \Delta t +  \sigma_2(p, \xi)\, \Delta W,
\end{aligned}
\end{equation}
where $p=p_{t_n}$ and $\xi=\xi_{t_n}$.
In this case the momentum $p_{t_{n+1}}$ is purely deterministic.
However, the pitch angle has a deterministic component, $b_2 \Delta t$, and a stochastic component given by the 
Brownian motion increments $\Delta W$ drawn form a Gaussian distribution,
$ \exp(-\Delta W^2/2 \Delta t)/\sqrt{2 \pi \Delta t}$,  with zero mean and standard deviation $\sqrt{\Delta t}$.
The expectation in Eq.~(\ref{e7}) can  be computed by sampling the Gaussian distribution.
However, as mentioned before, this direct method is not efficient because  the slow convergence rate of random sampling
requires a very large number of samples. 
The alternative method proposed here exploits the fact that (within the Euler step) the increments are Gaussian 
and Eq.~(\ref{e7}) can thus  be approximated as
\begin{equation}\label{e8}
\begin{aligned}
P(t_n, p, \xi) & \approx 
\frac{1}{\sqrt{2 \pi \Delta t}} \int_{\mathbb{R}} P\left(t_{n+1}, p+ b_1 \Delta t, \xi +b_2\Delta t +  \sigma_2 x\right)\; \exp\left({-\frac{1}{2} \frac{x^2 }{\Delta t}}\right) dx \, ,
\end{aligned}
\end{equation}
for small $\Delta t$.
The computation of  $P(t_n, p, \xi)$ knowing $P(t_{n+1}, p, \xi)$ is then reduced to the evaluation of the integral in  Eq.~(\ref{e8}) that can be 
efficiently computed with high accuracy using the  Gauss-Hermite quadrature rule according to which
\cite{Zhang2010, Zhang2012}
\begin{equation}\label{e9}
\begin{aligned}
P(t_n, p, \xi) & \approx  \sum_{m= 1}^M w_m P(t_{n+1}, p^{\rm GH}, \xi^{\rm GH}_m),
\end{aligned}
\end{equation}
where $b_1 = b_1(p, \xi)$, $b_2 = b_2(p,\xi)$, $ \sigma_2 =\sigma_2(p,\xi)$, $M$ is the number of quadrature points, $\{w_m\}_{m=1}^M$ are the set of weights and $\{\xi^{\rm GH}_m\}_{m=1}^M$ are the set of quadrature points, defined by
\begin{equation}\label{e12}
 \xi_m^{\rm GH} = \xi+b_2(p, \xi)\Delta t+ \sigma_2 (p,\xi)\sqrt{2\Delta t}\;q_m,
\end{equation}
for $m = 1, \ldots, M$ and $\{q_m\}_{m = 1}^M$ is the standard Gauss-Hermite abscissa that can be found in classic textbooks on numerical analysis (e.g., \cite{2013JSV...332.4403B}). The approximation error of the $M$-point quadrature rule is on the order of $\mathcal{O}\left(\frac{(\Delta t)^M M!}{(2^M (2M)!)}\right)$, so that $M = 3$ is sufficient 
to match the $\Delta t$ accuracy of the weak convergence of the Euler scheme in Eq.~\eqref{e6}.
That is, due to the $(\Delta t)^M$ factor in the quadrature error, $M$ does not have to be large to achieve convergence.

According to Eq.~(\ref{e6}) there is a finite, although very small,  probability that  within $[t_n, t_{n+1}]$, with $\Delta t \ll 1$, a realization of $\xi_t$ might lie outside $[-1,1]$. As a result there might be quadrature points $\xi_m^{\rm GH}$ that lay outside the domain $[-1,1]$. When this is the case, quadrature points outside the domain $[-1,1]$ are reflected back into $[-1,1]$. This prescription is mathematically equivalent to imposing a Neumann boundary condition. 

Equation~\eqref{e8} provides the value of $P$ at a fixed point $(p,\xi)$. To compute the $P$ in the phase space domain
$ [p_{\min}, p_{*}] \times [-1,1]$ we introduce an $N_p \times N_\xi$ uniform grid 
 \begin{equation}\label{e10}
 \mathcal{S} = \left\{ (p_i, \xi_j)| p_{i} = p_{\min}+i\Delta p, \,\xi_j = -1 + j\Delta \xi \right\}, \, 
 \end{equation}
 $i = 0, \ldots, N_p$,  $\Delta p = (p_{*}-p_{\min})/N_p$, $j = 0, \ldots, N_\xi$,  $\Delta \xi = 2/N_\xi$, 
and use Eq.~\eqref{e8} to compute  $P(t_n, p_i, \xi_j)$ at $(p_i, \xi_j) \in \mathcal{S}$ for each time step. 
For any point that is not on the grid $\mathcal{S}$, we approximate the solution $P(t_n,p,\xi)$ using piecewise linear interpolation, i.e.,
 \begin{equation}\label{e11}
 \tilde{P}(t_n,p, \xi) = \sum_{i = 0}^{N_p} \sum_{j = 0}^{N_\xi} c_{n, i, j}\, \varphi_i(p)\, \varphi_{j}(\xi),
 \end{equation}
where $c_{n,i,j} \approx  {P}(t_n,p_i, \xi_j)$ are the approximations based on Eq.~\eqref{e8} at the grid point $(p_i, \xi_j)$, 
 $\varphi_i(p)$ and $\varphi_j(\xi)$ are the linear finite element basis functions (i.e., the hat functions). Note that the interpolant $ \tilde{P}(t_{n+1},p, \xi)$ is used to evaluate the quadrature points (in Eq.~\eqref{e8}) that are not on the grid $\mathcal{S}$. It should also be noted that some quadrature point $(p^{\rm GH}, \xi^{\rm GH}_m)$ in Eq.~\eqref{e8} may locate outside the domain $[p_{\min}, p_{*}] \times [-1,1]$ in which $\tilde{P}(t_{n+1},p, \xi)$ is defined. In this case, we evaluate $ \tilde{P}(t_{n+1},p^{\rm GH}, \xi^{\rm GH}_m)$ using the following rules: 
 if $p^{\rm GH} > p_{*}$, $\tilde{P}(t_{n+1},p^{\rm GH}, \xi^{\rm GH}_m) = 1$; if $p^{\rm GH} < p_{*}$, $\tilde{P}(t_{n+1},p^{\rm GH}, \xi^{\rm GH}_m) = 0$; if $p^{\rm GH} \in [p_{\min}, p_{*}]$ and $\xi^{\rm GH}_m < -1$, $\tilde{P}(t_{n+1},p^{\rm GH}, \xi^{\rm GH}_m) = \tilde{P}(t_{n+1},p^{\rm GH}, \xi^{\rm GH}_m + 2(-1-\xi^{\rm GH}_m))$; if $p^{\rm GH} \in [p_{\min}, p_{*}]$ and $\xi^{\rm GH}_m > 1$, $\tilde{P}(t_{n+1},p^{\rm GH}, \xi^{\rm GH}_m) = \tilde{P}(t_{n+1},p^{\rm GH}, \xi^{\rm GH}_m - 2(\xi^{\rm GH}_m-1))$. This strategy is equivalent to adding a Dirichlet boundary condition on the boundary of $ [p_{\min}, p_{*}]$, and a no-flow Neumann boundary condition on the boundary of $[-1,1]$.
 Once $\tilde{P}(t_n, p, \xi)$ is computed for $n = 0, \ldots, N$, the runaway probability $P_{\rm RE}$ in Eq.~\eqref{REP} can be immediately obtained by substituting $\tilde{P}(t_n, p, \xi)$ into Eq.~\eqref{PPP}, i.e.,
 \[
 P_{\rm RE}(t_n, p, \xi) \approx \tilde{P}(T-t_n, p, \xi) \quad \text{ for }\; n = 0, \ldots, N.
 \]

The proposed implementation of the BMC method can be summarized in the following algorithm:

\vspace{0.5cm}
\begin{tabular}{p{0.95\textwidth}}
\hline\vspace{-0.35cm}
Algorithm 1\\
\hline
\vspace{-0.8cm}
\begin{algorithmic}\label{algorithm1}
\State {\bf Input}: $p_{\min}, p_{*}$, $Z$, ${E}$, ${\tau}$, $M$, $\Delta t$, $\Delta p$, $\Delta \xi$
\State {\bf Output}: The interpolant $\tilde{P}(t_n, p,\xi)$ for $n = N-1,\ldots, 0$.
\State Define the terminal condition $\chi(p, \xi)$ in Eq.~\eqref{chafun};
\State Generate the partition $\mathcal{T}$ for $[0,T]$ based on Eq.~\eqref{e9};
\State Generate the partition $\mathcal{S}$ for $[p_{\min}, p_{*}] \times [-1,1]$ based on Eq.~\eqref{e10};
\State Construct an interpolant $\tilde{P}(T_N, p, \xi)$ using \eqref{e11} by evaluating $\chi(p,\xi)$ at $\mathcal{S}$.
\For{$n = N-1, \ldots, 0$}
\For{$i = 0, \ldots, N_p, j = 0, \ldots, N_\xi$}
\State Generate the quadrature weights $\{w_m\}_{m=1}^M$
\State Generate the quadrature points $\{ \xi^{\rm GH}_m\}_{m=1}^M$ by plug $\xi_j$ into Eq.~\eqref{e12};
\State Evaluate the interpolant $\tilde{P}(t_{n+1}, p, \xi)$ at $\{ p^{\rm GH} , \xi^{\rm GH}_m\}_{m=1}^M$;
\State Compute $c_{n,i,j}$ using Eq.~\eqref{e8};
\EndFor 
\State Construct $\tilde{P}(t_n,p,\xi)$ by substituting $c_{n,i,j}$ into Eq.~\eqref{e11}.
\EndFor
\vspace{-0.7cm}
\end{algorithmic}\\
%
\hline
\noalign{\smallskip}
\vspace{-2cm}
\end{tabular}
\vspace{-1cm}

\begin{figure}[h!]
\center
\includegraphics[scale = 0.45]{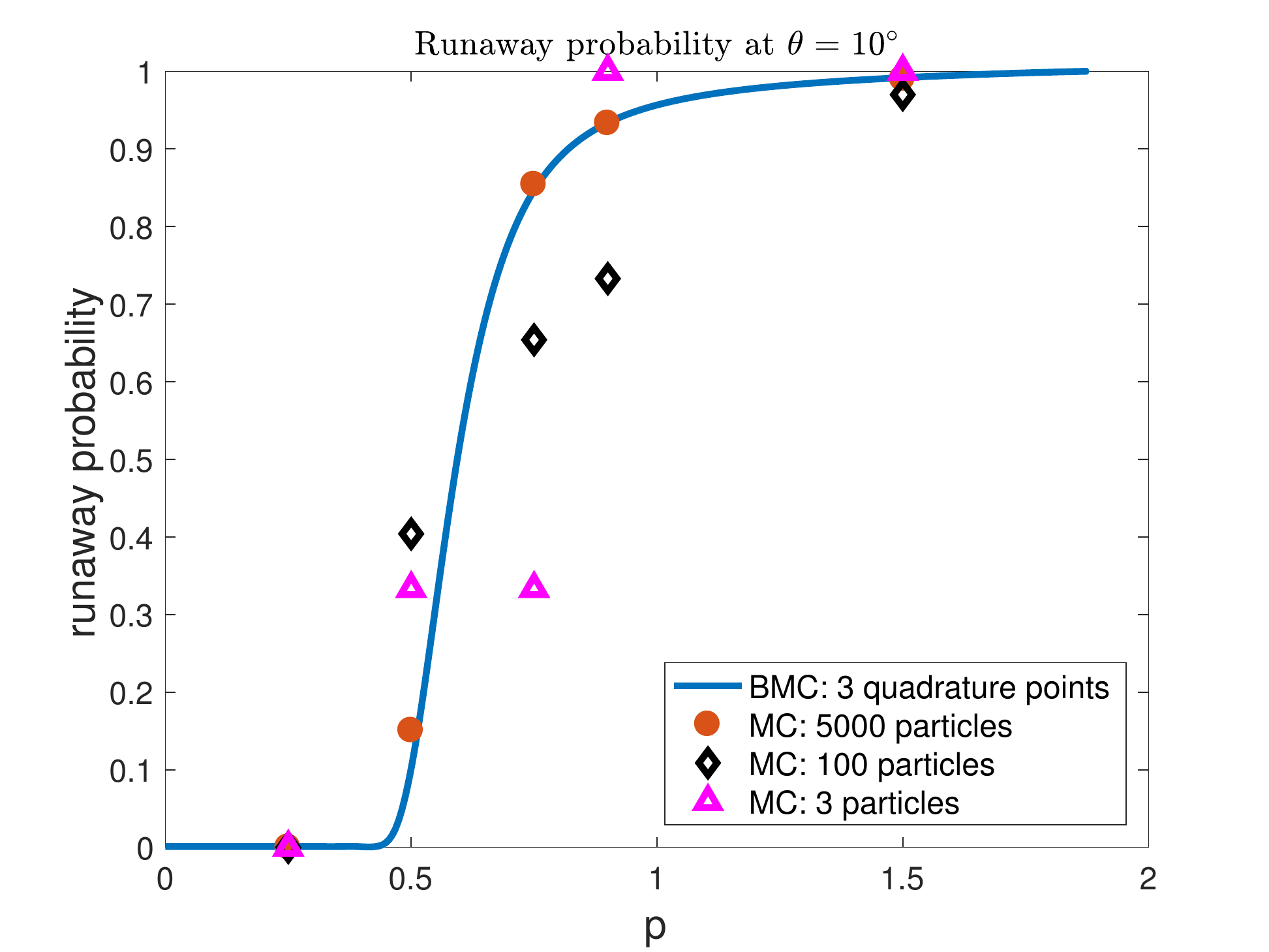}
\caption{Comparison between the BMC and the direct MC for pitch angle $\theta=10^{\circ}$
 and $T = 1.6$.}\label{fig_compare_MC-BMC}
\end{figure}

Figure~\ref{fig_compare_MC-BMC} compares $P_{\rm RE}$ computed using the BMC method with a direct MC simulation for 
pitch angle $\theta = 10^{\circ}$ and $T = 1.6$. Other parameters are set to $\tau = 1$, $E = 6$, $Z = 1$, $p_{\min} = 0$, $p_{*} = 2$ and $T/\Delta t = 2^8$. The computational complexities of the MC and BMC methods are proportional to the number of particles and the number of quadrature points, respectively. The idea is to compare the accuracy given the same complexity for both methods. To this end, we use 3 quadrature points  (i.e., setting $M = 3$ in Eq.~\eqref{e8}), and 3 particles for MC to simulate the SDEs in Eq.~\eqref{e1}. In this case, as expected, it is clear that the BMC method is much more accurate than the MC method. Even taking into account some extra cost of BMC in evaluating the interpolant in Eq.~\eqref{e10}, the BMC method is also more accurate than the MC with 100 particles, which is much more expensive than the BMC algorithm. The second aspect is to compare the complexity given the same accuracy for both methods. In this case, we use 5,000 particles for the MC simulation. The MC simulation can provide almost the same accuracy as the BMC method with significantly more amount of computational cost. 

\begin{figure}[h!]
\includegraphics[scale = 0.45]{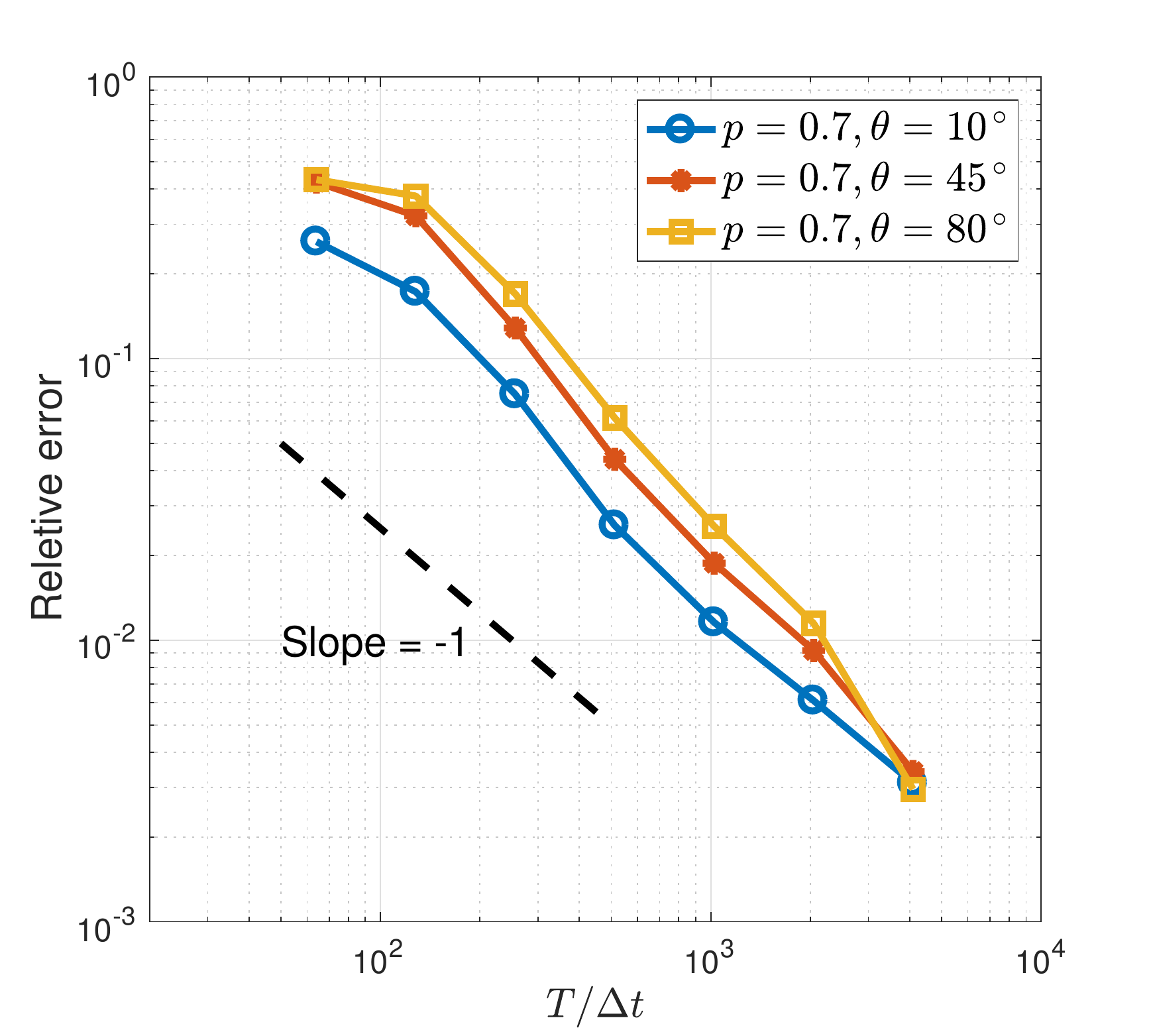}
\caption{Decay of relative error in the computation of $P_{\rm RE}(T, p, \theta)$ using the BMC method, at  $(p, \theta) = (0.7, 10^{\circ}),  (0.7, 45^{\circ}), (0.7, 80^{\circ})$. The dashed line shows a linear scaling of the error with $T/\Delta t$.} \label{fig10}
\end{figure}

To further illustrate the accuracy and efficiency of the BMC method, we conduct a benchmark example 
with  $\tau = 1$, $E = 6$, $Z = 1$, $p_{\min} = 0$, $p_{*} = 2$, $T = 8$ in Eq.~\eqref{KBE}.
The reference runaway probability is obtained using the classic forward Monte Carlo method with $10^8$ realizations of the SDE in Eq.~\eqref{e1}. To test the convergence of the BMC method we choose $T/\Delta t = 2^{6}, 2^{7}, 2^{8}, 2^{9}, 2^{10}, 2^{11}, 2^{12}$ and set $\Delta t = \Delta p = \Delta \xi$. Figure \ref{fig10} shows the decay of the error in approximating the reference $P_{\rm RE}(T, p, \theta)$ at three selected points in the phase space, i.e., $(p, \theta) = (0.7, 10^{\circ}),  (0.7, 45^{\circ}), (0.7, 80^{\circ})$. We can see that our approach can achieve a first-order convergence with respect to $\Delta t$, which is more accurate than the forward MC method. Moreover, the BMC method can recover the entire profile of $P_{\rm RE}(t, p, \xi)$ for all possible initial condition $(p, \xi)$, while the forward MC has to resample the SDE in Eq.~\eqref{e1} for different initial conditions. 

\begin{figure}[h!]
\center
\includegraphics[scale = 0.45]{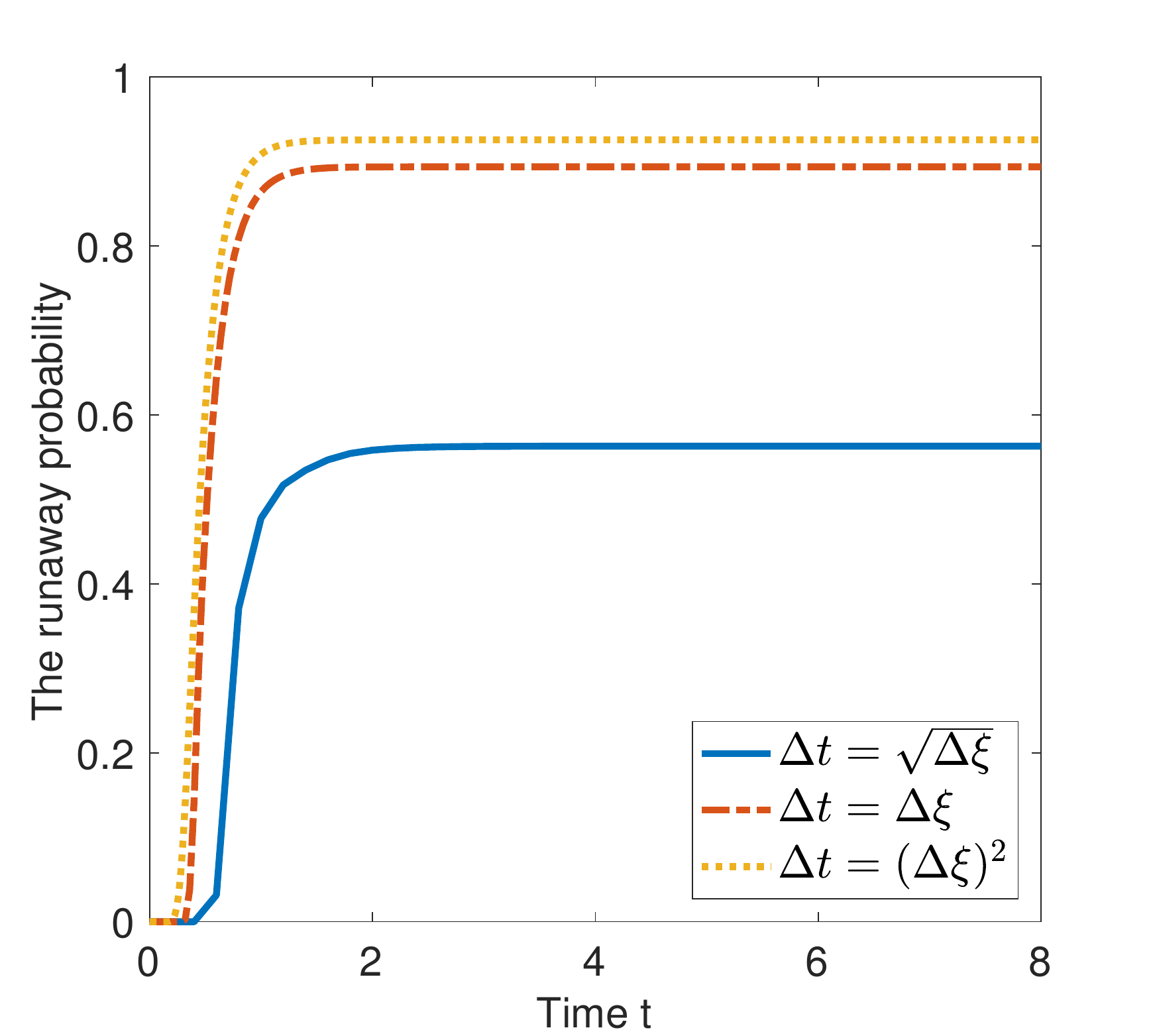}
\caption{Illustration of stability of the BMC method. For a fixed momentum space grid size, $\Delta p$, and widely different time steps, $\Delta t$, the computation of the runaway probability is stable. In particular, the different curves show  that in all the cases 
$P_{\rm RE}(t,p,\cos \theta)$ at  $(p,\theta) = (1, 10^{\circ})$ remains finite.}\label{fig11}
\end{figure}

The BMC method also presents advantages over solving the adjoint partial differential equation (PDE) in Eq.~\eqref{KBE} using finite different/finite element or spectral  methods. 
First, as Eq.~\eqref{e7} shows, in the BMC  the time-stepping scheme is {\em explicit}, and thus there is no need to solve a linear system at each time step. On the other hand, the use of explicit schemes for the adjoint PDE usually leads to severe instability, unless $\Delta t$ is  small enough. However, as shown in Ref.~\cite{Zhang2012}, even though an explicit scheme is used, the BMC method is {\it absolutely stable} for any $\Delta t$. To illustrate this, Fig.~\ref{fig11} shows the evolution of the approximate runaway probability $P_{\rm RE}(t,p,\cos \theta)$ based on the BMC method at  $(p,\theta) = (1, 10^{\circ})$. Here, we fix $\Delta p = \Delta \xi = 0.04$ and investigate three cases $\Delta t = \sqrt{\Delta p} =  0.2$, $\Delta t = \Delta p =  0.04$ and $\Delta t = (\Delta p)^2 =  0.0016$. Despite different approximation errors, the BMC method is stable in all the cases,
regardless of the value of $\Delta t$. 

Due to the use of Gauss-Hermite quadrature rule and Lagrange/Hermite interpolation, our scheme can achieve comparable convergence rates as classic PDE approaches, like $\mathcal{O}(\Delta t)$ in time discretization (using implicit Euler scheme), and $\mathcal{O}((\Delta p)^4+(\Delta \xi)^4)$ in phase space (using piecewise cubic interpolation).  Adaptive time-stepping algorithms for partial differential equations can be used to achive accuracy higher than 
$\mathcal{O}(\Delta t)$. In this regard, we should mention that going beyond the low order Euler step used here,
the evolution of the deterministic (non-stochastic) part of the dynamics could be done using a higher order, possible adaptive, time stepping algorithm. 
Further details, and rigorous error analysis of the proposed BMC method can be found in Refs.~\cite{Zhang2010,Zhang2012,Zhang2013,Zhao_etal_2017}.

The BMC can be easily extended to solve higher dimensional problems. When the additional degrees of freedom do not involve diffusive transport the extension is straightforward and it does not incur in a significant computational overhead. On the other hand, adding degrees of freedom with diffusive dynamics requires computing a high-dimensional version of the quadrature formula in Eq.~(\ref{e8}). For example, 
a more accurate collision operator  would introduce energy diffusion in the model in Eq.~(\ref{FP}) and thus
add a stochastic component to the momentum evolution in Eq.~(\ref{e6}), 
\begin{equation}\label{e6b}
\begin{aligned}
p_{t_{n+1}} & = p + b_1(p, \xi)\, \Delta t,  + \sigma_1(p, \xi)\, \Delta W,\\
\end{aligned}
\end{equation}
where $\sigma_1$ denotes the momentum diffusivity. 
In this case, Eq.~(\ref{e7}) is approximated by the double-integral 
\begin{equation}\label{e8b}
\begin{aligned}
P(t_n, p, \xi) & \approx 
\frac{1}{2 \pi \Delta t} \int_{\mathbb{R}^2} P\left(t_{n+1}, p+ b_1\Delta t+ \sigma_1 y , \xi +b_2\Delta t +  \sigma_2 x\right)\; 
\exp\left({-\frac{x^2+y^2 }{2\Delta t}}\right) d x d y\, ,
\end{aligned}
\end{equation}
which, as before, can be efficiently and accurately computed using  tensor-product Gauss-Hermite quadrature rules  with $M^2 = 9$ quadrature points. It should be noted that Eq.~\eqref{e8b} will not change the representation of the interpolating polynomial in Eq.~\eqref{e11}. However, when the SDE system in Eq.~\eqref{e1} involves more than two dynamical processes   the total computational cost will increase. A case of particular interest is the addition of a radial spatial degree of freedom, which will yield a $3$-D dynamical system capable of modeling spatial confinement. Going beyond this, we envision applying this method to $5$-D guiding center and $6$-D full orbit models of runaway electrons.  In these cases, 
the use of sparse grid techniques 
 \cite{Zhang2013}  will be required to avoid the exponential growth of computational cost. In particular, if the maximum mesh size in each direction is $h$, a $d$-dimensional cartesian grid will have a total of $\mathcal{O}((1/h)^d)$ grid points. In comparison, a sparse grid only needs $\mathcal{O}((1/h) (\log(1/h))^{d-1})$ grid points, which is much smaller than $\mathcal{O}((1/h)^d)$.

%
%
%
%
%
%
%

\section{Applications}
\label{applications}
Here we illustrate the versatility of the BMC method in the computation of several statistcal observables of physical interest in the study of runaway electrons.

\subsection{Time dependent and asymptotic probability of runaway $P_{\rm RE}$.}
\begin{figure}[h!]
\center
\includegraphics[scale = 0.5]{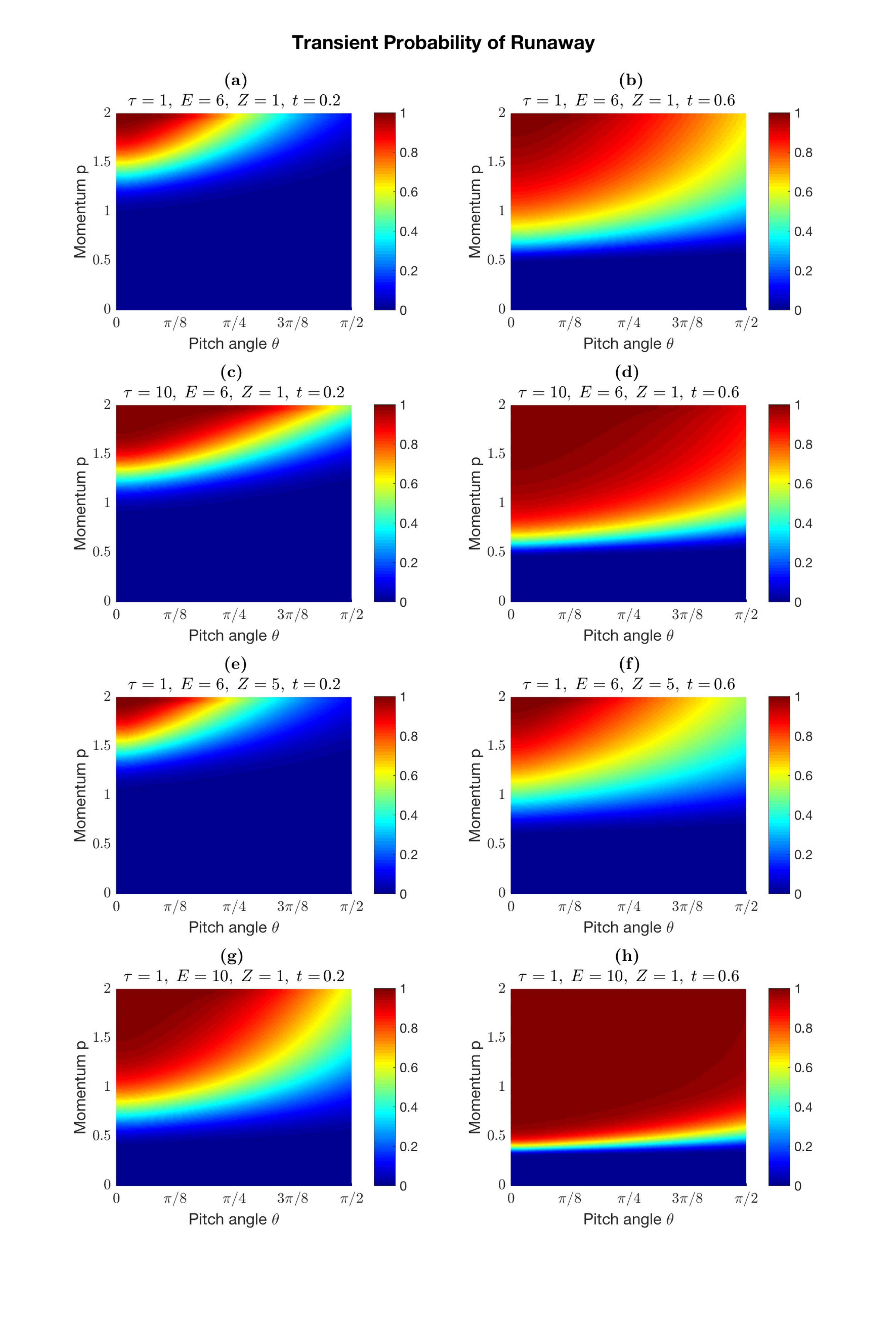}
\caption{Transient probability of runaway electrons at time $t = 0.2$ (left column), and $t=0.6$ (right column) for different values of $\tau$, $E$ and $Z$.
}\label{fig1}
\end{figure}

As discussed in Sec.~\ref{model}, the definition of $P_{\rm RE}$ depends on  $p_*$, which is the momentum determining the 
runaway boundary. In most of the calculations that follow we will assume $p_*=2$. However, to test the dependence of the results on $p_*$ we will also consider $p_*=6$. 
Figure~\ref{fig1} shows the time evolution of $P_{\rm RE}$ for $p_*=2$ in the $(p,\theta)$ momentum space for different values of $\tau$, $E$, and $Z$,
which, as discussed in Sec.~\ref{model}, are free parameter of the Fokker-Planck model.
The left and right columns show the corresponding snapshots of $P_{\rm RE}$ at $t=0.2$ and $t=0.6$ respectively. 
Panels (a) and (b) show the ``reference case", $(\tau,E,Z)=(1,6,1)$. As panels (c) and (d) show, for fixed $E$ and $Z$, as $\tau$
increases, the runaway region (i.e. the region in phase space for which $P_{\rm RE} \sim 1$) grows. This is consistent with the fact that, as  Eq.~(\ref{rr})
indicates, an increase in $\tau$ implies a decrease in the radiation reaction force and thus a higher runway acceleration. 
On the other hand,  as shown in panels (e) and (f), for fixed $E$ and $\tau$,   the runaway region shrinks as $Z$ increases, a results consistent with 
Eq.~(\ref{col}) that shows that a higher $Z$ leads to a larger pitch angle scattering and thus larger energy losses due to an increase on synchrotron emission.  Finally, as expected, as panels (g) and (h) show, for fixed $\tau$ and $Z$, the runaway region grows as $E$ increases. 

In the limit $t \rightarrow \infty$, $P_{\rm RE}$ reaches a steady state. Numerically it is observed that for the parameters under study, for  $t > 8$,
$P_{\rm RE}$ does not change significantly and thus, for practical purposes we refer to $P_{\rm RE} (t=8, p,\theta)$ as the time-asymptotic, steady-state runway probability distribution. The left column in Fig.~\ref{fig2} shows this distribution for the same parameter values as those in 
Fig.~\ref{fig1}. The dependence of the asymptotic runaway region on the parameters follows the pattern discussed above
for the time dependent runaway region, namely,  it increases with $\tau$ and $E$ and decreases with $Z$. 
The solid black lines in the plots on the left column of Fig.~\ref{fig2} show the predictions based on the test particle model. 
This simplified model, originally proposed and studied in Refs.~\cite{fuchs_etal_1986,solis_etal_1998}, neglects the role of collisions and thus reduces the stochastic differential equations in Eq.~(\ref{e1}) to the deterministic dynamical system  $dp/dt=b_1(p,\xi)$, $d\xi/dt=b_2(p,\xi)$. This dynamical system has an hyperbolic fixed point, and the corresponding stable manifold provides a separatrix that corresponds to the boundary of the runaway region in phase space. Since this model  neglects pitch angle scattering, it provides a sharp, step-function type, approxiamte boundary. However, as expected  this boundary is close to the $P_{\rm RE}=0.50$ contour line, i.e. the boundary determining the initial conditions that with will runaway with a $50 \%$ probability.   

\subsection{Expected runaway time $T_{\rm RE}$ and loss time $T_{\rm Loss}$.}

\begin{figure}[h!]
\center
\includegraphics[scale = 0.47]{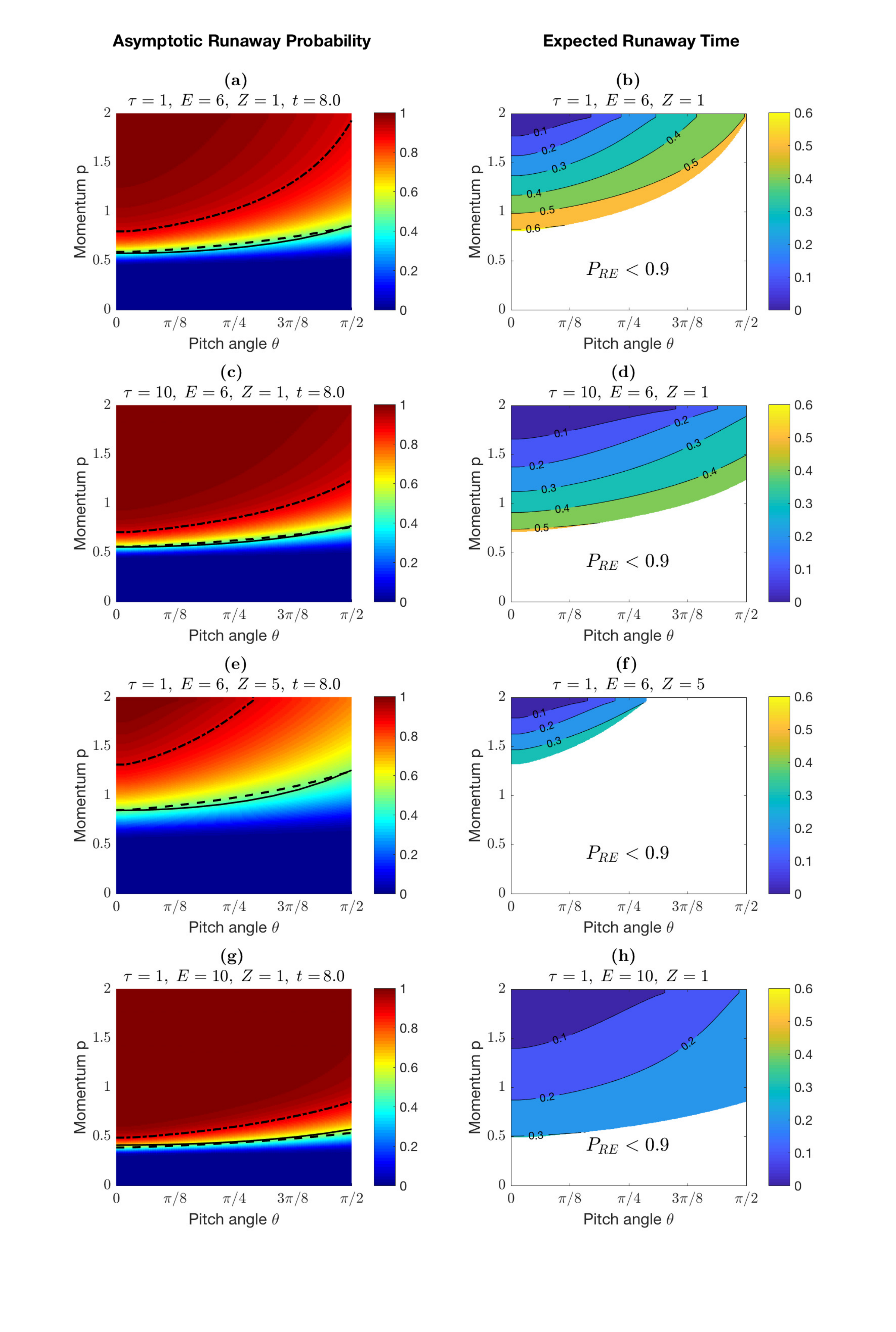}
\caption{Left column: Time asymptotic runaway probability where the dash-dot line ``-\,$\cdot$\,-\,$\cdot$'' is the $P_{\rm RE} = 0.9$ contour and the dashed line ``- - -'' the $P_{\rm RE} = 0.5$ contour. The solid line denotes the test particle model (without diffusion) prediction. 
Right column: Expected runaway time, $T_{\rm RE}$, for $P_{\rm RE} \ge 0.9$.}\label{fig2}
\end{figure}

Another useful observable is the expected runaway time, $T_{\rm RE}$, defined as
 \begin{equation}\label{e20}
 T_{\rm RE}(p,\xi) = \frac{1}{P_{\rm RE}(\infty,p,\xi)}\int_0^{\infty} t \frac{\partial P_{\rm RE}(t,p,\xi)}{\partial t} dt \, .
 \end{equation}
The computation of $T_{\rm RE}$ in the  phase space needs to be restricted to $(\theta,p)$ initial conditions that have a fixed, given probability of runaway. As an example, the right column of Fig.~\ref{fig2} shows  $T_{\rm RE}$ in the phase space region for which the probability of runaway is larger than $90 \%$ ($P_{\rm RE}=0.9$ ). As expected, $T_{\rm RE}$ is maximum along the  $P_{\rm RE}=0.9$ contour and it decreases as $\theta$ decreases and $p$ increases. Formally, the white regions in these plots correspond to 
$T_{\rm RE} = \infty$.  

\begin{figure}[h!]
\center
\includegraphics[scale = 0.47]{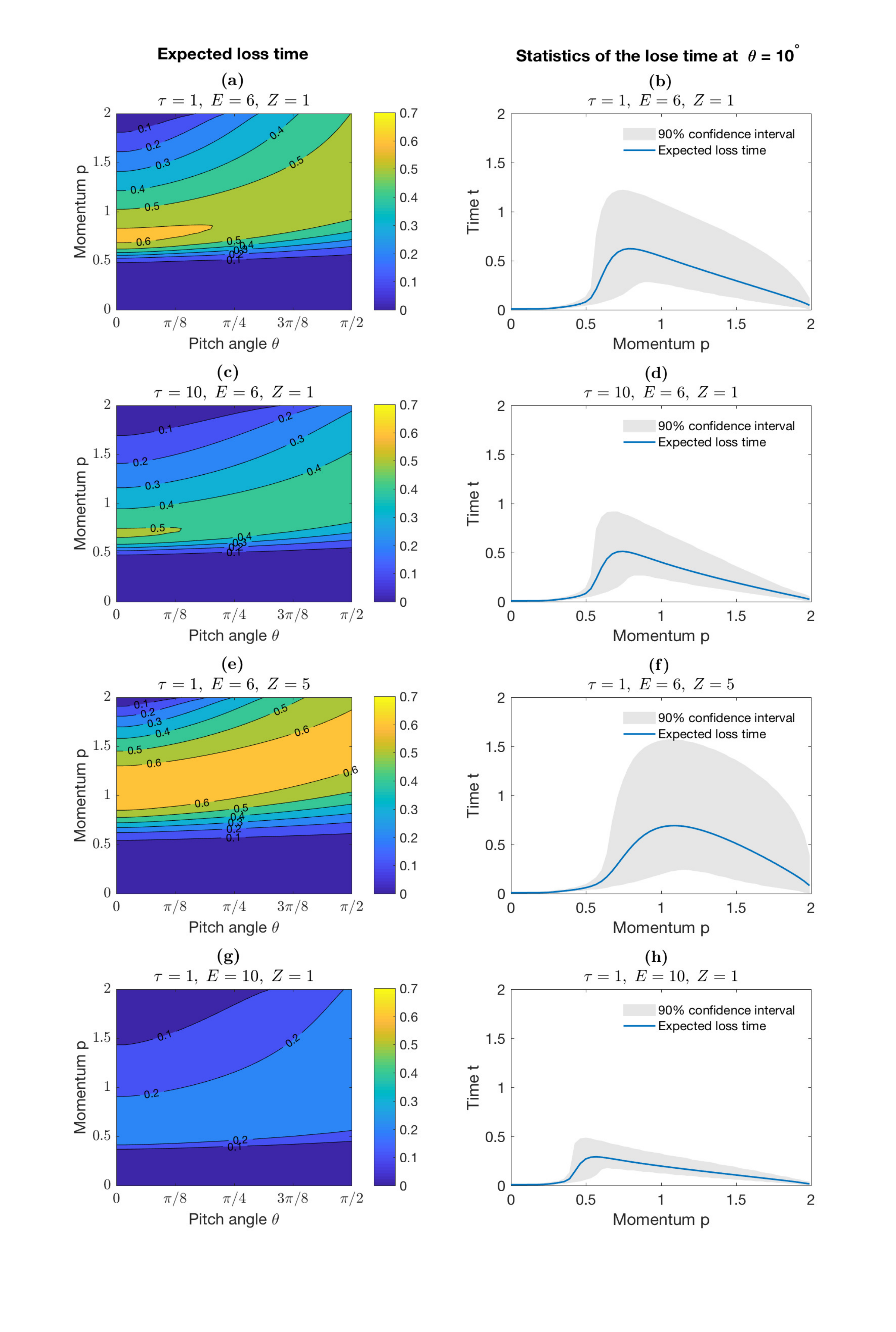}
\caption{Left column: expected loss time, $T_{\rm Loss}$;
Right column: Confidence intervals of the loss time.
}\label{fig3}
\end{figure}

To compute the expected runaway loss time, $T_{\rm Loss}$, we need to calculate the evolution of $P_{\rm Loss}$ that gives the
 probability of an electron hitting either the high energy bound (i.e., $p_{*}$) or the low energy bound (i.e., $p_{\min}$).
We obtain $P_{\rm Loss}$ by solving Eq.~\eqref{chafun} using the BMC  Algorithm 1 with 
terminal condition 
\begin{equation}\label{chafun1}
\chi_{\rm Loss}(p_t,\xi_t) = 
\left\{
\begin{aligned}
&1, \;\; \text{ if } \; p_t \ge p_{*} \text{ or } \; p_t \le p_{\min},\\
&0, \;\; \text{ otherwise} \, .
\end{aligned}
\right.
\end{equation}
Knowing $P_{\rm Loss}$, $T_{\rm Loss}$ is computed using Eq.~\eqref{e20} with  $P_{\rm RE}$  replaced by $P_{\rm Loss}$.
Figure \ref{fig3} shows $T_{\rm Loss}$ for the four cases considered in Fig.~\ref{fig2} and $p_*=2$. 
The right hand column in this figure shows the 90\% confidence interval of the loss time for pitch angle $\theta = 10^{\circ}$. 
The size of the confidence interval describes the extent of uncertainty of the loss time. The bigger the confidence interval, the more uncertainty in the loss time. 

\begin{figure}[h!]
\center
\includegraphics[scale = 0.71]{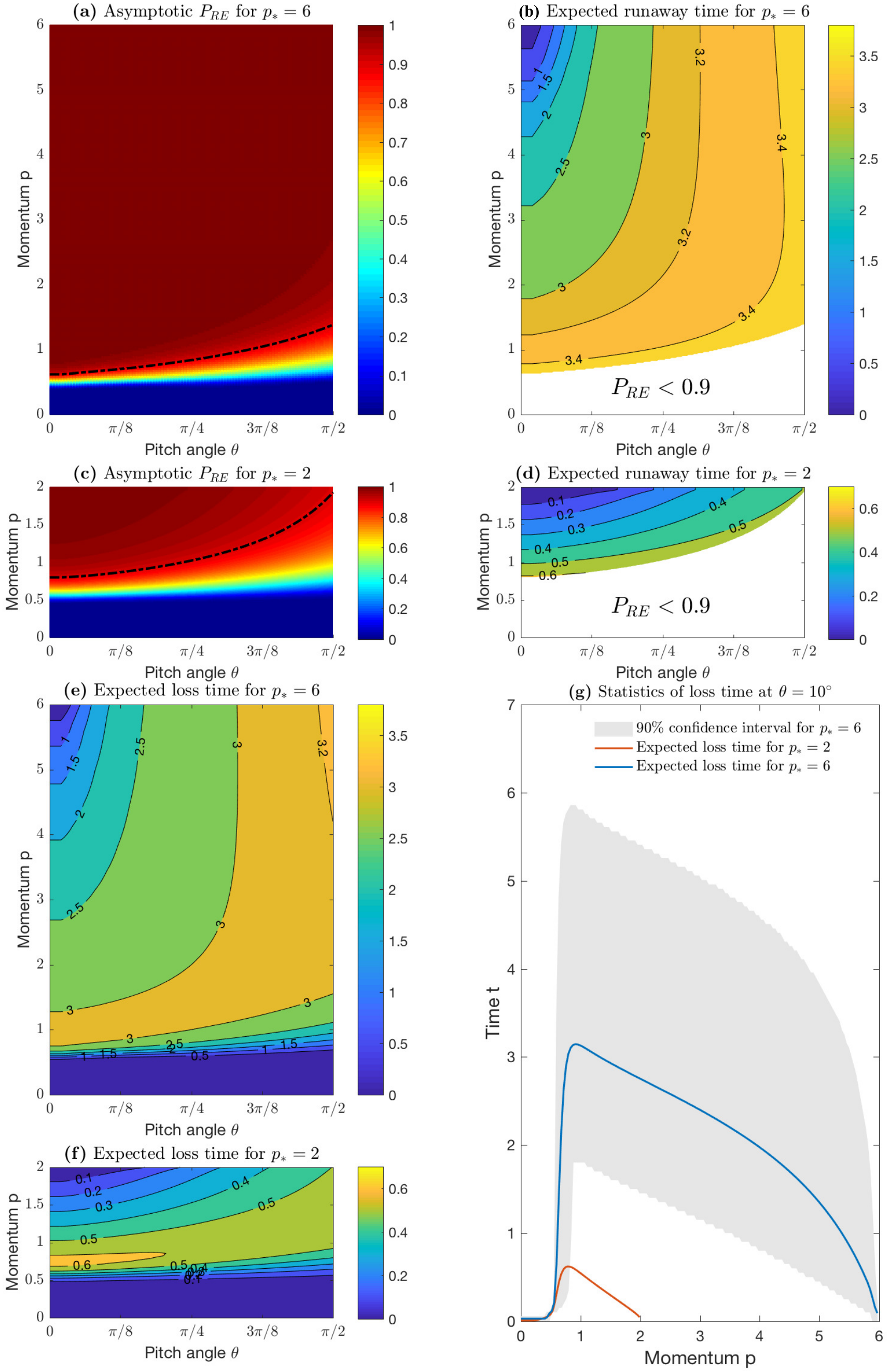}
\caption{(a) Asymptotic $P_{\rm RE}$ for $p_{*} = 6$; (b) Expected runaway time for $p_{*} = 6$; (c) Asymptotic $P_{\rm RE}$ for $p_{*} = 2$; (d) Expected runaway time for $p_{*} = 2$; (e) The expected loss time for $p_{*} = 6$; (f) Expected loss time for $p_{*} = 6$; (g) Comparison of  mean and 90\% confidence interval of loss time for $\theta = 10^{\circ}$}\label{fig4}
\end{figure}

To explore the dependence on $p_*$ of the previous results, Fig.~\ref{fig4} shows $P_{\rm RE}$, $T_{\rm RE}$ and $T_{\rm Loss}$ for $p_*=6$. 
By comparing panels (a) and (c) in Figure \ref{fig4}, the first observation is that setting $p_{*} = 6$ does not dramatically change the runaway probability distribution for the domain $p \in [0,2]$; the second observation is that runaway probability in the region $p \in [2,6]$ are very close to 1, which demonstrate the rationality of the terminal condition used in adjoint equation \eqref{KBE}. By comparing the panels (b),(e) to (d),(f), we observe that expected runaway and loss times for the case of $p_{*} = 6$ is longer than the case of $p_{*} = 2$. It is reasonable because it will take longer for a particle to hit either the high energy or the low energy bounds when the domain becomes bigger.
 In Figure \ref{fig4}(g), we plot 90\% confidence interval of the loss time for the case $p_{*} = 6$. We can see that the confidence interval is bigger than that of the case $p_{*} = 2$, which means that the derivative of the runaway probability $P_{\rm RE}$ (with respect to time $t$) in the case $p_{*}=6$ is smaller than that in the case of $p_{*} = 2$. This phenomenon is also observed in 
 the production rate in Figure \ref{fig_pr_t}. The bigger $p_{*}$, the slower the production rate grows. 

\subsection{Production rate}

\begin{figure}[h!]
\center
\includegraphics[scale = 0.45]{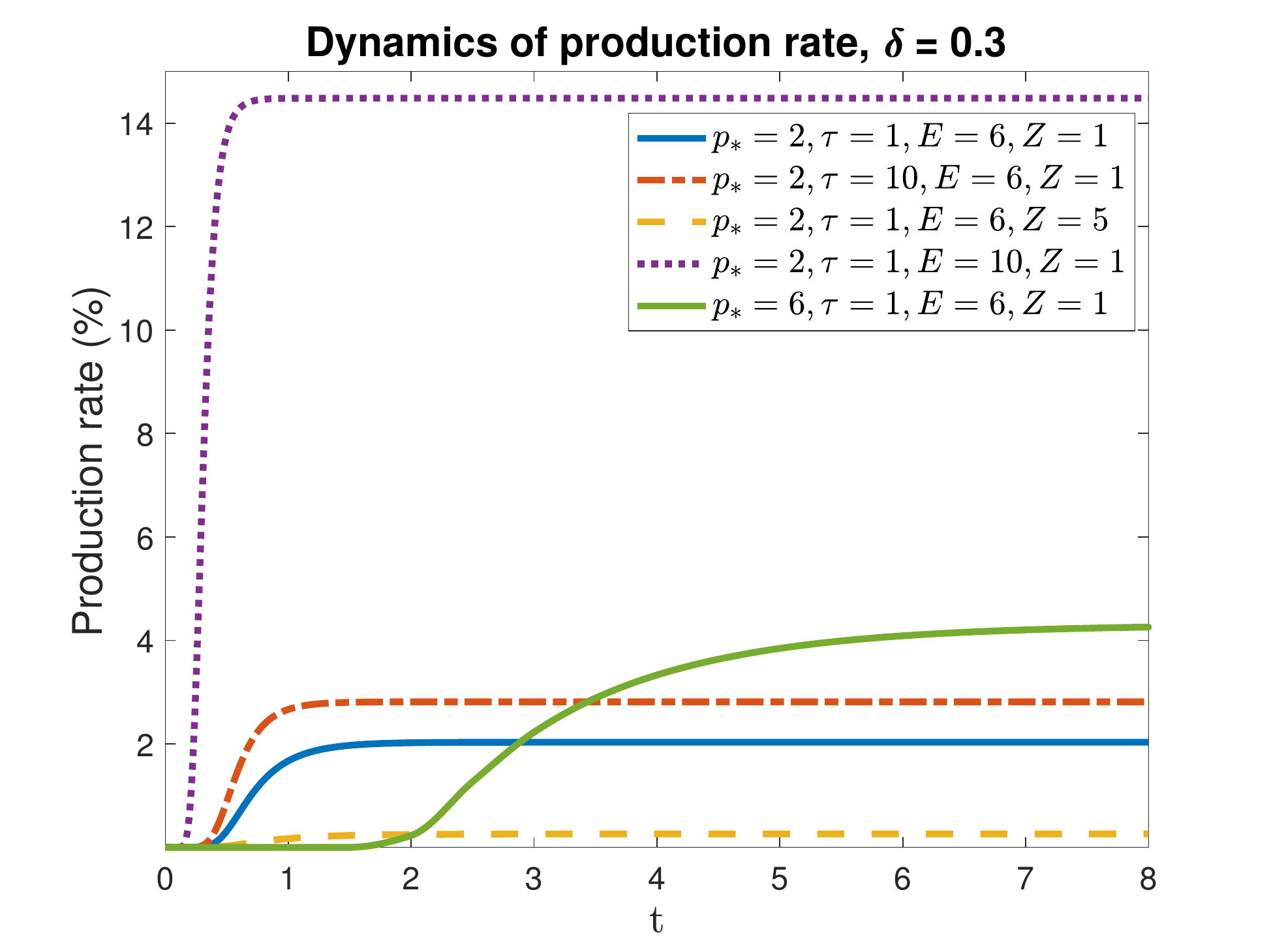}
\caption{Time evolution of production rate of runaway electrons according to Eq.~(\ref{prate}) for different values of $\tau$, $E$ and $Z$.}
\label{fig_pr_t}
\end{figure}

\begin{figure}[h!]
\center
\includegraphics[scale = 0.45]{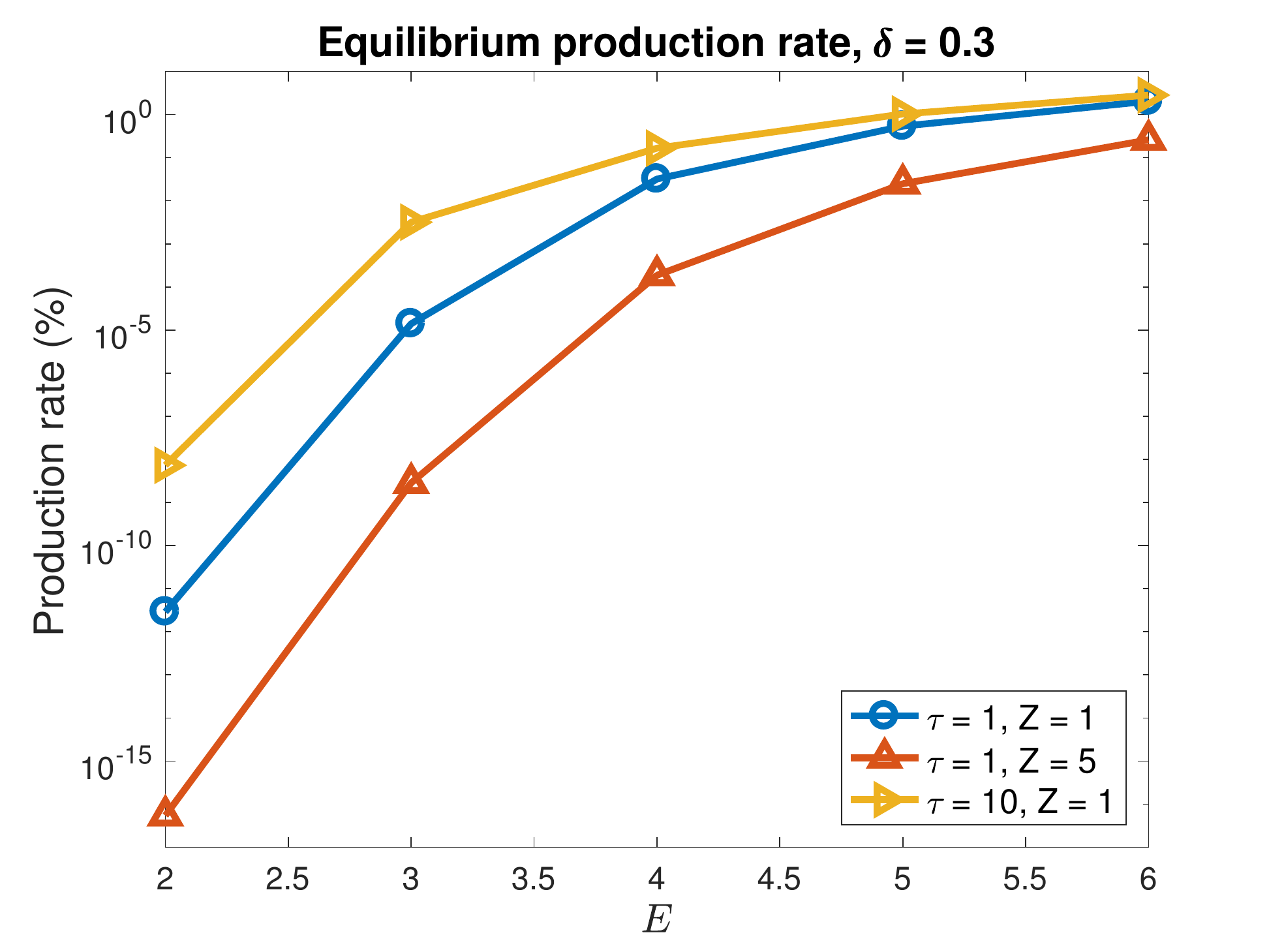}
\caption{Time asymptotic, equilibrium value of production rate of runaway electrons according to Eq.~(\ref{prate}) as function of $E$, for different values of $\tau$ and $Z$.}
\label{fig_pr_e}
\end{figure}

To conclude we apply the BMC method to compute the runaway electron production rate, 
\bq
\gamma=\frac{N_{\rm RE}(t)}{N} =  \int_0^\infty  dp  \int_{-1}^1 d \xi \,  f(p, \xi) P_{\rm RE} (t,p,\xi) \, ,
\eq
where $N_{\rm RE}(t)$ is the number of RE at time $t$, $N$ is the total number of electrons, and $f$ is the distribution function of the thermal electrons. 
For a Maxwellian distribution, 
\begin{equation}
f_M(p,\xi)=\frac{2p^2}{\pi^{1/2} \delta^3} e^{-\left(p/\delta\right)^2} \, ,
\end{equation}
normalized as 
$\int_{-1}^1\int_{0}^{+\infty}f_M(p,\xi) dp d\xi = 1$ 
we have 
\bq
\label{prate}
\gamma(t)= \frac{2}{\sqrt{\pi} \delta^3} \int_0^{p_{*}} dp \,  e^{-\left(p/\delta\right)^2} p^2  \int_{-1}^1 d \xi \,  P_{\rm RE} (t,p,\xi)\, 
 + \gamma_{\infty} \, ,
\eq
where we have explicitly introduced the upper bound $p_*$ for which, by definition, $P_{\rm RE} (t,p>p_*,\xi)=1$, and 
%
\bq
\gamma_{\infty}=\left[\frac{1}{2} {\rm erfc}\left( \frac{p_{*}}{\delta} \right)+\frac{1}{\sqrt{\pi}} \left( \frac{p_{*}}{\delta} \right) e^{-\left( \frac{p_{*}}{\delta} \right)^2} \right] \, .
\eq
Figure~\ref{fig_pr_t} shows $\gamma(t)$ according to Eq.~(\ref{prate}) for $p_*=2$ and $6$, with $\delta=0.3$. It is observed that for fixed $E$ and $Z$, the production rate 
increases with $\tau$. This is to be expected because an increase in $\tau$ implies a reduction of synchrotron radiation damping. 
On the other hand, for $\tau$ and $E$ fixed, as $Z$ increases the production rate significantly decreases because the increase on 
pitch angle scattering due to collisions leads to an increase of synchrotron radiation losses.   This same trends are observed in the 
asymptotic, equilibrium value of the production rate that shows an expected monotonically increasing dependence on the electric field $E$, as shown in Figure \ref{fig_pr_e}. 

 \section{Summary and Conclusions}
 \label{conclusions}
We have proposed a novel backward Monte Carlo method for an accurate and efficient  computation of the time-dependent probability of runaway. The method is based on the direct numerical solution of the Feynmann-Kac formula.   
Starting from the final runaway state at $t=T$, the method reconstructs iteratively the probability of runaway at previous times $t=T-\Delta t, \, T- 2\Delta t, \, \ldots$, where $\Delta t \ll 1$ is the step size. At each time step
the algorithm reduces to the computation of an integral  involving the previously computed probability of runaway and the Gaussian propagator.
This integral is efficiently computed with high accuracy using the  Gauss-Hermite quadrature rule. Points outside the computational phase-space grid are evaluated using pice-wise linear interpolation.  

As shown by the numerical simulations, the proposed method achieves the advantages of both the PDE method (i.e., high-order accuracy) and the ``brute-force'' Monte Carlo method (i.e., easy parallelization), as well as overcomes their disadvantages. 
In particular, the method is unconditionally stable and it requires very few quadrature points to achieve high accuracy. 
 
It is important to remark that the reason we name our algorithm ``Backward Monte Carlo'' is because it is based on a probabilistic representation, i.e., conditional expectations with respect to the underlying SDEs. However, we do not use random sampling of paths to compute the conditional expectations. 
Instead, we use Gauss-Hermite quadrature rules, which actually makes the proposed algorithm deterministic. 

To illustrate the versatility of the method we have presented a study of the time evolution and asymptotic steady state  of the probability of runaway as function the relative strength of the synchrotron radiation reaction, $1/\tau$,  the ion effective charge, $Z$, and the magnitude of the electric field $E$. We also have computed the expected runaway time, the expected loss time, and the 
production rate which play a key role in  the understanding of the  dynamics of runaway electrons. 

Even though the applications of the BMC method focused on a simplified 2D RE model, it is straightforward to extend it to high-dimensional cases by exploiting sparse approximation techniques, e.g., sparse-grid interpolation \cite{Zhang2013}. These extensions will be presented in a future publication. 


 \section{Acknowledgments}
 
We thank E. Hirvijoki for valuable comments and suggestions. 
This material is based upon work supported in part by the U.S.~Department of Energy, Office of Science, Offices of Advanced Scientific Computing Research and Fusion Energy Science
; and by the Laboratory Directed Research and Development program at the Oak Ridge National Laboratory, which is operated by UT-Battelle, LLC., for the U.S.~Department of Energy under Contract DE-AC05-00OR22725.

\end{document}